\documentclass[a4paper]{article}
\usepackage{graphicx}
          \begin{document}

          \title{The self-organised phases of manganites}

          \author{N.D. Mathur \\ Department of Materials Science,
University of Cambridge, CB2 3QZ, UK\\
           and \\P.B. Littlewood \\ Department of Physics, University of
Cambridge, CB3 0HE, UK}

          \date{\today}
          \maketitle

          \begin{abstract}

Self-organisation requires a multi-component system. In turn, a
multi-component system requires that there exist conditions in which more
than one component is robust enough to survive. This is the case in the
manganites because the free energies of surprisingly dissimilar competing
states can be similar --- even in continuous systems that are chemically
homogeneous. Here we describe the basic physics of the manganites and the
nature of the competing phases. Using Landau theory we speculate on the
exotic textures that may be created on a mesoscopic length scale of
several unit cells.

          \end{abstract}

          \section{Introduction}

The doped perovskite manganites are fascinating because they can readily
be tuned between radically different phase states. Broadly speaking one
has either a ferromagnetic metal, a charge-ordered insulator, or a
paramagnetic polaron liquid. The spatial extent of these three types of
phase is potentially unlimited --- but so, according to a growing body of
evidence, are the number of alternative phases that occur over mesoscopic
length scales.

What is the basis for such richness and complexity? The answer is that the
magnetic, electronic and crystal structures of any given manganite are
intimately related. Thus one may parameterise any given phase in the
manganites by the nature of the spin, charge, orbital and structural
degrees of freedom; the degree to which these degrees of freedom are
static or dynamic; and, of particular interest here, the length scales
over which these degrees of freedom are homogeneous. Note that we do not
consider any texture that only extends over microscopic (i.e. unit cell)
length scales to be a phase. For example, one stripe in a stripe phase is
not in itself a phase --- because at least one dimension is too small for
the object to be considered a thermodynamic entity.

At a microscopic level one can understand the strong interaction between
the magnetic, electronic and crystal structures as follows. The electronic
sub-lattice of a perovskite manganite consists of a cubic network of
corner-sharing $MnO_{6}$ octahedra in which potential charge carriers
arise in certain orbitals if, for example, the interstitial A-site cations
comprise a mixture of trivalent and divalent species such that they act as
a charge reservoir. This electronic sub-lattice is also the home of the
magnetic sub-lattice since the magnetic structure arises primarily from
the magnetic nature of the manganese atoms. Therefore the electronic and
magnetic sub-lattices are one and the same such that they must interact
strongly. The intimate connection with the crystal lattice arises both
because $Mn$ is Jahn-Teller (JT)  distorted by charge carriers, and also
because the radius of the A-site cations is invariably less than ideal. 

Several recent reviews provide colourful insights into the macroscopic and
microscopic nature of the manganites \cite{reviews, elPSrevs}. Here we
concentrate on sub-micron {\em mesoscopic} length scales over which there
is growing evidence for multi-phase coexistence. We argue that Landau
theory will provide a natural explanation for these mesoscopic phenomena,
and suggest a strategy that would enable self-organised structures to be
generated in a controlled manner.

As a caveat we note that historically, evidence for phase separation in
the manganites has come from several directions; an old reference known to
us dates back to 1977 \cite{Leung}. The evidence we describe in this
article is by no means complete. For example, we concentrate on the cubic
manganites and neglect the Ruddlesden-Popper layered manganites in which
phase separation has been seen \cite{Argyriou}.

          \section{Basic physics of manganites}

\begin{figure}
          \centerline{\includegraphics[height=2in]{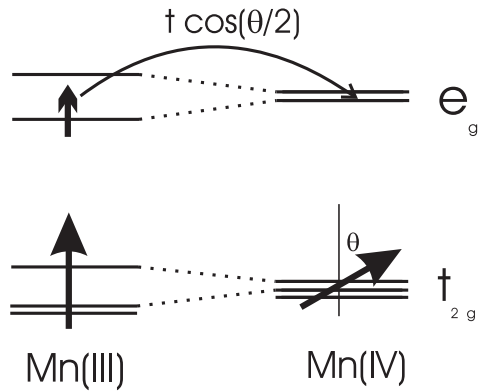}}
          \caption{\label{Fig1} Electronic crystal field levels of
$Mn^{3+}$ and
          $Mn^{4+}$ ions in a crystal. The $Mn^{3+}$ is shown in the
presence of a
Jahn-Teller splitting
          of the $e_g$ (and $t_{2g}$) orbitals, due to breaking of the
cubic
symmetry
caused by a
          distortion of the oxygen octahedron. The $Mn^{4+}$ is shown with
the cubic symmetry
          preserved; since the upper level is unoccupied, the energy
cannot be lowered by
          a Jahn-Teller distortion.
          }
          \end{figure}

          $La_{1-x}Sr_xMnO_3$ is a prototype for the broad class of cubic
perovskite manganites, where with replacement
          of a trivalent rare earth by a divalent alkaline earth, the
{\em nominal} valence of $Mn$ can be continuously tuned
          between $3+$ (corresponding to the Mn(III) configuration found
at x=0) and $4+$ (corresponding to the Mn(IV) configuration found at x=1).
          The important physical ingredients can be seen by reference to
Fig. \ref{Fig1}.

In a cubic environment, the $Mn$ $d$ levels are crystal-field split into a
low lying triplet of $t_{2g}$ symmetry and a doublet of $e_g$ symmetry. 
$Mn$ is a strongly correlated ion whereby double occupancy of the tightly
bound $d$ orbitals is suppressed by Coulomb repulsion, and the direct
on-site exchange interaction aligns the spins in the different $d$
orbitals. The $t_{2g}$ levels are strongly localised, whereas an electron
in the higher lying $e_g$ state is potentially itinerant. At $x=0$, each
$e_g$ level is singly occupied: double occupancy of these highest occupied
$e_g$ levels is suppressed by Coulomb repulsion, and $LaMnO_3$ is an
antiferromagnetic Mott insulator. At finite doping ($x > 0$), there are
some empty $e_g$ levels, and hence hopping is possible. The exchange
(Hund's rule) coupling $J$
          between the spin of an itinerant carrier and each core
spin is rather larger than the hopping matrix element $t$ between
          neighbouring $e_g$-levels. Consequently, each conduction
electron is forced to align with the core spin texture (this may be viewed
as a strong-coupling version of the RKKY interaction). Thus the
          effective hopping matrix element between $Mn$ sites is $t
\cos(\theta/2)$, where $\theta$ is the relative angle between neighbouring
          core spins. Clearly, the kinetic energy of the conduction
electrons is minimised (maximum bandwidth) if the core spins are
          parallel to one another, and this so-called {\em double
exchange} is the
fundamental
mechanism \cite{double_exchange, deGennes} of metallic ferromagnetism at
low temperatures in the doped manganites. At a sufficiently high
temperature the energy of this ferromagnetic configuration is overwhelmed
by the entropy gain available from a randomisation of the manganese spin
system.  Thus the system lowers its free energy by entering the {\em
paramagnetic} state.

          Another important feature of the manganites arises from the
doubly-degenerate $e_g$ level in Fig. 1. This degeneracy may be
          broken by a Jahn-Teller (JT) distortion of the oxygen cage away
from cubic symmetry that lowers the energy of the occupied $e_g$ level
          on the $Mn^{3+}$ ion. In pure $LaMnO_3$, there is an
antiferrodistortive arrangement of the distorted octahedra
\cite{kanamori},
          leading to a ``doubled'' unit cell. This is to be distinguished
from an equivalent and simultaneous source of doubling of the unit cell
due to rigid rotations of the octahedra, which are promoted by the small
size of the A-site cations. (These rotations mainly affect O-Mn-O bond
angles whereas the Jahn-Teller distortions mainly affect O-Mn-O bond
lengths.) Since the $e_g$ level is progressively
          depopulated with increasing $x$, the tendency toward JT
distortion is suppressed; the long-range-ordered antiferrodistortive
          phase disappears near $x=0.2$. It is important to note that the
lattice displacements associated with the JT distortions are large
          and therefore the disappearance of the long-range order does not
mean that (static or dynamic) JT fluctuations may not
          be pronounced. Such fluctuations have been discovered to be very
prominent in the manganites.

On top of these JT fluctuations --- be they static or dynamic --- are the
additional static distortions mentioned earlier that arise because the
radii of the A-site cations are invariably less than the 1.30 ${\AA}$
required for the ideal manganese perovskite structure \cite{Attfield}.
Looking through the Attfield chemical window \cite{Attfield} one can see,
for example, that at an electronic doping level that is particularly
favourable for ferromagnetism (x=0.3), a high Curie temperature is
achieved when the average A-site cation radius is large and the variance
in this quantity is small --- such that distortions and disorder are
respectively minimised. 

Both these JT fluctuations, and small A-site cations of varying radii,
promote mixed-valent {\em insulating} phases of the manganites. The Mn
(III) configuration will have a tendency to distort, whereas the Mn(IV) 
configuration will not. Because every electron in a local $e_g$ level
carries around its own JT lattice distortion, this would appear to provide
a recipe for insulating phases at any composition. 

\subsection{Low temperature ferromagnetic metallic phases (FMM)}

If one is close to the ideal perovskite structure in the (x=0.3) Attfield
chemical
window then enough of the $O-Mn-O$ bonds are sufficiently linear to make
$t$ so large that the ground state is ferromagnetic by the means described
above --- whence the charge carriers are delocalised such that JT
fluctuations are suppressed. One of these delocalised charge carriers is
like the quantum-mechanical ``particle in a box'': its energy is lowered
when the box is made bigger, provided that any internal structure within
the box can be ignored.

\subsection{Charge-ordered insulating phases (COI)}

Even when the JT fluctuations do indeed make for an insulating ground
state, there can be order because the oxygen octahedra share corners;
distortions about one $Mn$ site are anti-shared with a neighbour.
Steric conflicts can be solved in ordered phases (such as undoped
$LaMnO_3$), and in TEM it is common to observe ``striped'' phases
consisting of
diagonal arrays of Mn(III) and Mn(IV)  \cite{stripes}. These striped
phases predominate in $x\ge$0.5 where the larger and asymmetric Mn(III)
species are in the minority such that packing is easier. In these
localised phases, the spin couplings are via superexchange, mostly
antiferromagnetic, and naturally somewhat smaller than the ferromagnetic
double exchange interactions in the metal.

In fact the term COI is used here to refer to a multiplicity of phases,
including the parent antiferrodistortive ($x$=0) phase, and the
          striped phases that have been reported with several different
periods that are commensurate with the lattice, especially periods of 2
(near $x$=1/2) and 3 (near $x$=2/3) \cite{stripes}. A naive counting of
the ratio of Mn(III)/Mn(IV) would predict
          {\em incommensurate} phases with either one Mn(IV) stripe every
$(1/x)$ in a Mn(III) environment (0$<x\le$0.5), or one Mn(III) stripe
every $1/(1-x)$ in a Mn(IV) environment (0.5$\le x<$1). However, there
          are undoubtedly strong commensurability effects that will
stabilise the commensurate phases away from their
          ``correct'' nominal composition. The nature of the repeating
units in stripe phases --- be they commensurate or not --- is not
universally agreed upon. In the prevailing ``bi-stripes'' picture
\cite{stripes} the widths of the Mn(III) and Mn(IV) stripes are different.
In the Wigner crystal model they are of equal width \cite{Wigner}. There
are also phases that may be charge-ordered, but not orbitally ordered ---
i.e. the orientation of the JT distortions are random, or strongly
fluctuating locally
 \cite{Radcondmat}. We shall suggest some other possibilities later. 

\subsection{The high temperature paramagnetic phase (PMI)}

The recent burst of interest in the manganites was initially motivated by
the discovery reported in 1994 of the bulk ``colossal'' magnetoresistive
(CMR) effect in the
          {\em paramagnetic} phase at temperatures just around the Curie
point \cite{Jin}. In situations and compositions (e.g.
$La_{0.7}Ca_{0.3}MnO_3$) where
a large CMR is achieved in this way, the paramagnetic
          state is found to be quasi-insulating (activated resistivity),
and this has been tracked to the presence of large
          polaronic (JT) fluctuations in the paramagnetic insulator
(PMI) which are strongly suppressed in the
          FMM \cite{polaron_theory, neutPDFs, deTeresa, Billinge_recent,
EXAFS}. In this case, the high temperature competition for the FMM is not
with the ordered mixed-valent insulator,
          but with something resembling a melted version of it --- let us
call this a ``polaronic liquid''.

Note that nearer to the ideal perovskite structure in the (x=0.3) Attfield
chemical window the dynamic polaronic fluctuations and hence CMR are
suppressed in the paramagnetic phase. This is the case with
$La_{0.7}Sr_{0.3}MnO_3$ above $T_C$ --- which consequently has a metallic
resistivity in that it rises with increasing temperature \cite{LSMO}. In
fact, there is no symmetry distinction between this paramagnetic metal
(PMM) and the PMI, and the crossover (with varying composition) between
these phases is generally smooth.

\section{Competing phases}

The three basic phases described above (COI, FMM and PMI) compete to be
the stable thermodynamic phase. What can happen near the FMM/COI boundary
is of particular interest in this article because there is no reason of
symmetry why charge ordering and ferromagnetism cannot coexist. Instead it
depends on the driving physics. This suggests that the transition between
the two types of ordered phases will be first order --- which seems to be
the case (at least at low temperatures).

Below we argue that the spatial extent of a ``phase'' cannot be too small,
and go on to describe some of the evidence for texture in the manganites
that arises over length scales that are at least microscopic. We then
consider the strong evidence for texture over longer length scales, i.e.
``phase separation''.

\subsection{Microscopic fluctuations}

The PMM/FMM transition in $La_{0.7}Sr_{0.3}MnO_3$ described in Section 2.3
is continuous (second order) and smooth at the microscopic level as
evidenced by ferromagnetic resonance measurements \cite{FMR}. The FMM/PMI
transition (e.g. in $La_{0.7}Ca_{0.3}MnO_3$ as also described in Section
2.3) is also usually continuous \cite{bulkev}. However more generally, a
{\em thermodynamically} continuous (second order) transition is not
necessarily smooth microscopically: it could be an order-disorder
transition rather than a microscopically smooth ``displacive'' order-order
transition.  This order-disorder type of continuous (second order) 
transition can only be distinguished from a {\em thermodynamically}
discontinuous (first order)  transition if the length scale of the
measuring probe is sufficiently long.  Experimentally there is plenty of
evidence to demonstrate the existence of at least {\em short-range}
texture at the unit cell level, some of which we now describe below.

NMR studies in an applied magnetic field can probe both the ionisation
state and the {\em local} magnetic environment of the $^{55}Mn$ atoms. All
around the phase diagram it is typical to find Mn(IV)/Mn(III) species
coexisting with Mn species whose NMR signal is motionally narrowed
\cite{NMRrev}. Such motional narrowing is indicative of the presence of
metallic regions. However, in principle, motional narrowing can also arise
if the energy levels accessible to a somewhat localised electron become
blurred (e.g. due to thermal or interaction effects). Therefore, to be
pedantic, we consider the existence of a motionally narrowed line to be
good evidence rather than concrete proof for the existence of metallic
phases. This technicality notwithstanding, NMR studies lead us to believe
that there are metallic clusters in surprising places: above $T_C$ on the
$x<0.5$ hole-doped side \cite{NMRclusters} and at 3 K on the $x>0.5$
electron-doped side where AF stripe phases predominate \cite{NMRPRL}. A
particularly interesting feature of NMR is that one can see whether or not
the frequency of a motionally narrowed resonance will fall with the bulk
magnetisation on increasing the temperature \cite{NMRrev, Savosta}: if it
does fall then we have a second order displacive transition; if it does
not then we infer that the $Mn$ atoms are not all sitting in the same
environment. However, since NMR is sensitive only to the local
environment, this does not establish the existence of two {\em
thermodynamic} phases: the disorder might only arise at the microscopic
level.

Neutron data can reveal short-range order (as well as long-range order,
discussed later). For example, in the $La_{1-x}Ca_{x}MnO_3$ system when
$x=0.3$, recent zero-field neutron spin echo studies of the low
temperature FM phase have demonstrated inhomogeneity on a length scale of
${\le 30 \AA}$; and parallel muon studies suggest that the spin dynamics
in the different regions are very different \cite{muons}. Moreover,
inelastic neutron scattering experiments have revealed that this low
temperature FM phase contains soft zone boundary magnons and phonons
\cite{ZBmagnons}. 

Short-range order in $x=0.3$ systems has also been seen near $T_C$ using
neutrons: the spin wave stiffness is found to remain finite
\cite{Lynn,Adams} (c.f. the NMR evidence described above); and a
superlattice with correlation length ${10 \AA}$ is found to develop
\cite{Adams}. This latter finding extends earlier studies near $T_C$ in
which neutron pair distribution functions demonstrated the existence of JT
distortions with short-range correlations \cite{neutPDFs}. Similarly,
early findings with SANS revealed the existence of local magnetic clusters
with a length scale of ${12 \AA}$ \cite{deTeresa}. There is similar
evidence for these short-range correlations near $T_C$ from a recent high
energy x-ray powder analysis \cite{Billinge_recent}; and EXAFS studies in
which the $Mn$ K-edge resonance is affected by local correlations
\cite{EXAFS}.

Regarding short-range order at other doping densities: in the
$La_{1-x}Ca_{x}MnO_3$ system at $x=0.15$, the FMM ground state found at
higher $x$ is replaced with a novel form of short-range orbital and charge
order, according to bond valence sum calculations based on neutron powder
diffraction data \cite{Lobanov}.  It is around this doping density that
canted spins \cite{deGennes} were originally postulated --- effectively as
an interpolation between the antiferromagnetic end member at $x=0$ and the
FMM region at higher $x$.  The evidence for novel short-range order
presented in \cite{Lobanov}, and the presence of both ``FM'' and ``AFM''
peaks in NMR studies using an applied magnetic field \cite{Allodi}, tend
to rule out canting in favour of the postulated \cite{Lobanov} microscopic
texture. However the two scenarios are not mutually exclusive and in
$LaMnO_{3.02}$ the 250 MHz NMR line displays a field-dependent slope that
corresponds to half a gyromagnetic ratio; this indicates a canted state
\cite{NMRrev}. 

In $x\ge0.5$ there is both neutron and TEM evidence for the so-called
stripe phases that constitute the COI phase. We discuss this region of the
phase diagram in the next section.

The experiments described above, and certain others, can be adduced as
evidence for the existence in the manganites of inhomogeneities that arise
over length scales that are comparable with unit cell dimensions. In some
cases such experiments point to, but do not prove, the existence of
inhomogeneities arising over somewhat longer length scales. These larger
inhomogeneities form the subject of this review and in the next section we
consider some of the strong and recent evidence for their existence. For
now let us be clear: we reserve the term ``phase separation'' for the
coexistence of mesoscopic and macroscopic thermodynamic phases. 

\subsection{Phase separation and coexistence}

Whenever there is a first order phase boundary, there can be coexistence
between phases. An unconstrained system in which there is a first order
phase boundary should in principle be in one phase or the other at
equilibrium. However in practice coexistence is liable to arise for
several reasons.

Equilibrium
          coexistence of two thermodynamic phases generally occurs either
because of disorder (where locally one phase is favoured
          over the other) or because of an imposed conservation law. The
classic case of the latter is of the liquid/gas transition in a system
          of fixed volume, where the equilibrium fractions of the liquid
and gas will adjust in order to fill the available space. When
          external parameters of the system are changed (e.g. temperature,
volume), the two phases readjust in different proportions.
          If the readjustment is impeded, there will be hysteresis.

The phenomenon of hysteresis is often cited as evidence for phase
coexistence. To be precise, it constitutes indicative but not conclusive
evidence. For example, hysteresis loops of magnetisation versus applied
magnetic field can arise in some rare-earth magnetic systems solely
because the degree of saturation dictates the strength of the reverse
field required to induce nucleation. In another example, hysteresis can be
produced by supercooling, but this is a kinetic phenomenenon.

          Electronic phase separation is often discussed in the context of
the manganites \cite{elPSrevs, el_phase_sep}. While it is no doubt a valid
          point of view to consider the mixed-valent charge-ordered phases
as examples of microscopic electronic phase separation, the Coulomb energy
          associated with substantial charge density fluctuations on
length scales longer than a few lattice constants is prohibitive.
          When we have coexistence between phases on long length scales,
the electronic charge densities must be very nearly the same. We stress
that in this and the subsequent sections of this article, we concern
ourselves with {\em phases} that by definition (see section 3.1) extend
over at least several unit cells.

A number of experiments have demonstrated that over wide ranges of
compositions there can be coexistence between the FMM and COI phases
(particularly period 2 or ``1/2''-type). Perhaps the most visual and
direct evidence for this type of coexistence comes from dark-field TEM
images of $La_{5/8-y}Pr_{y}Ca_{3/8}MnO_3$ at 20 K \cite{Uehara}.
Sub-micron domains appear bright because they contain the extra
periodicity associated with a COI phase, whereas the interspersed dark
regions of similar size are inferred to be FMM through percolative
resistivity measurements \cite{Uehara, Babushkina}. Neutron powder data
taken for compounds of similar compositions reveal the presence of both FM
and AFM peaks, each with a different temperature dependence
\cite{Balagurov}. This demonstrates directly the presence of two phase
coexistence: each phase must extend over at least some hundreds of
Angstroms in order to produce the neutron diffraction peaks. A more recent
analysis estimates this distance to be not less than 500 ${\AA}$ and
possibly even 1000 ${\AA}$ \cite{Sheptyakov}. A slightly more complicated
phase separation arises hysteretically in $Nd_{0.5}Sr_{0.5}MnO_3$ powder
samples below 145-175 K: the majority of the orthorhombic crystal, part of
which is a FMM, undergoes a transition to a monoclinic COI phase, as
evidenced by neutron diffraction and strain gauge data \cite{Ritter}.
Since a structural phase transition accompanies the appearance of the COI
phase, one anticipates the eventuality of strain-driven phase separation
in the manganites (see Section 4).

Further demonstrations of the FMM-COI interplay are found in sintered
powder/thin film experiments. Sintered samples of the potentially COI
material $La_{0.5}Ca_{0.5}MnO_3$ only develop a conducting path and a
large magnetisation when the grain size is small, i.e. when there is a
large amount of strained grain boundary material present \cite{Levy}.
Similarly, a strained epitaxial film of $La_{0.5}Ca_{0.5}MnO_3$ was found
to posess a FMM ground state \cite{Nyeanchi}, but this strain must be of
the correct type \cite{Chapman}. These experiments demonstrate the
important role of strain, which is expanded upon below and in the next
sections on Landau theory.
 
In the vicinity of the FMM/PMI phase boundary, Jaime {\em et al.}
\cite{Salamon} argued some time ago that transport and thermodynamic data
could be quantitatively described by a model of coexisting phases, rather
than a homogeneous transition. There is also some very visual imaging
evidence to support the strain-induced coexistence of the FMM and PMI
phases in manganites.  Scanning tunnelling microscopy scans suggest that a
thin strained film of $La_{0.7}Ca_{0.3}MnO_3$ contains both metallic and
insulating regions;  and that the insulating sub-micron regions even
persist at low temperatures and in high magnetic fields \cite{Fath}.
Alternatively, temperature dependent magnetic force microscopy studies
near an artificial grain boundary in a thin strained film of
$La_{0.7}Sr_{0.3}MnO_3$ reveal that the FMM state persists within a micron
of the boundary at temperatures above which the ferromagnetism in the rest
of the film survives \cite{Soh}. Again, the most likely interpretation of
these two experiments is that strain and strain relief respectively create
the long-range inhomogeneity.

Mesoscale coexistence of the COI phase with an insulating spin glass phase
over (500-2000) ${\AA}$ has been observed in a powdered sample of
$Pr_{0.7}Ca_{0.3}MnO_3$ using neutrons \cite{Radcondmat}. This coexistence
may be understood in terms of intragranular strain-driven segregation. The
glassy phase melts into a metal under the influence of an applied magnetic
field (c.f. melting the COI in $Nd_{0.5}Sr_{0.5}MnO_3$ \cite{Ritter}); and
in a single crystal of similar composition $(Pr_{0.63}Ca_{0.37}MnO_3)$
there is found to be a large latent heat associated with a similar
transition (i.e. thermodynamic evidence from calorimetry for first order
behaviour) and a time-dependent specific heat in the coexistence regime
that relaxes on a scale of tens of seconds \cite{Raychaudhuri}. Relaxation
phenomena have been observed elsewhere too, for example in ultra-thin
strain films of $La_{0.7}Ca_{0.3}MnO_3$ \cite{Blamire} and in
$Nd_{0.5}Ca_{0.5}Mn_{1-y}Cr_{y}O_3$ ${(0 \le y \le 0.1)}$ crystals
\cite{relaxorFM}.

Various other measurements of hysteresis \cite{oldHwang, Babushkina},
noise \cite{noise} and thermopower \cite{thermop} also indicate two-phase
coexistence; and various strategies for obtaining two-phase coexistence
have been employed. For example, on the Mn(IV)-rich side of the
$La_{1-x}Ca_{x}MnO_3$ phase diagram near $x=0.9$ a cluster glass state
develops and colossal magnetoresistance effects are seen \cite{Martin}. 
Similarly replacing 15\% of the $Mn$ atoms with $Cu$ to produce
$La_{2/3}Ca_{1/3}Mn_{0.85}Cu_{0.15}O_3$ appears to generate both FMM and
COI regions as evidenced by transport and magnetisation measurements as a
function of field and temperature \cite{MnCu}. 

Regarding the interpretation of the above examples (and many more), it is
clear that the properties of a manganite are liable to be strongly
dependent on the nature of the sample: the strain states and defects will
vary between single crystals, powders, sintered powders and thin films. We
note that from some perspectives, the propensity for twinning in the
manganites reduces the value of single crystal studies in favour of free
(non-sintered) powder samples since it is possible for each micron-sized
grain to act like an untwinned single crystal. One is confident that the
phase separation seen in neutron powder diffraction studies is not due to
surface effects since, as mentioned five paragraphs ago, each phase must
extend for hundreds of Angstroms to be seen at all with neutrons. Twinning
may also be avoided by recourse to thin films. 

Remarkably, the proportions of two coexisting phases can be easily
``tweaked'' by external parameters that include pressure \cite{pressure},
magnetic fields \cite{fields}, temperature \cite{Babushkina, oldHwang},
these three parameters together \cite{DonPaul}, strain \cite{Nyeanchi},
electrical currents \cite{currents} and X-ray \cite{xray} or electron
illumination \cite{Chenprivcomm}.

          We have argued elsewhere \cite{littlewood} that the
          {\em equilibrium} coexistence of two phases over a wide range of
parameter space is most likely produced
          by the clamping effects of external and internally-generated
strain fields. We must now ask what sets
          the length scale of the coexistence. We shall see that there is
an intriguing possibility that the competing phases can self-organise over
mesoscopic distances. Moreover, these
          self-organised structures may themselves constitute true
thermodynamic phases.

          \section{Landau theory}

It would be desirable, if possible, to construct a complete theory for the
manganites from a fully microscopic basis. In fact, there is a good
microscopic understanding of the homogeneous phases \cite{reviews,
elPSrevs}, and some evidence for texture near first order transitions
\cite{moreo}. A full understanding of mesoscopic texture from a
microscopic basis is a complex exercise
          because of the vast phase space to be explored. Moreover, such
an exercise could turn out to be unnecessarily reductionist. Instead then,
it may be useful, at least initially, to simplify the analysis by
          postulating a Landau theory for the various coupled order
parameters (magnetism, charge-order, strain). Such an approach has
          turned out to be very valuable in the case of
ferroelectric/ferroelastic transitions \cite{ferroelectrics},
          and especially valuable for the case of incommensurate and
commensurate phase charge-density wave systems \cite{mcmillan, CDW}. A
full Landau theory for the case at hand is still very complicated and has
not
          yet been fully explored \cite{brey}. We shall
          restrict ourselves to a few simple scenarios.  Of course, even
with its simplifications, the
          Landau theory can only demonstrate possibilities, because the
real physics lies in the values of the
          parameters of the theory, which we can only either guess at, or
fit to experiment.

          \subsection{Strain effects in a commensurate striped phase}
          In a phase transition to a striped periodic phase, there is a
development of an order parameter with a wavevector
          ${\bf Q}_i$, where the subscript denotes one of the (several)
directions of the possible stripes. Then we may write the
          full order parameter as
          \begin{equation}
          \rho_i f({\bf Q}_i \cdot {\bf r})
          \end{equation}
          so that $\rho_i$ is the amplitude, and the function $f$ is a
periodic function of its argument. The simplest Landau theory for the
          transition would then be contained in the following expansion of
the free energy
          \begin{equation}
          \label{frho}
          {\cal F_{\rho}} = \sum_i \left( \frac{1}{2} a(T) \rho_i^2 +
\frac{1}{4} b \rho_i^4 + \frac{1}{6} c \rho_i^6 + ...
          \right) + \sum_{i > j} b_{ij} \frac{1}{4} \rho_i^2 \rho_j^2 +
...
          \end{equation}
          For diagonal stripes, the sum must run over the possible
equivalent $\{110\}$ directions (we index throughout with respect to the
pseudo-cubic unit cell). For simplification
          we shall just consider the case when we may have one of the two
$(110)$ (labelled $i=1$, say) and
          $1\bar10$ $(i=2)$ directions in two dimensional symmetry.

          It is conventional to assume that $a(T) = a_o \times (T-T_o)$,
so that the quadratic coefficient changes sign below
          a temperature $T_o$. Depending on the sign of the coefficient
$b$, the model will predict a first order ($b<0$) or
          second order ($b>0$) transition. We shall need to choose $b_{12}
> 0$, so as to have a striped, rather than checkerboard
          phase as the lowest energy state.

          Because the presence of the order parameters breaks the cubic
symmetry, the order parameters must couple to shear strain $s$.
          This will lead to terms of the form
          \begin{equation}
          \label{fs}
          {\cal F}_{s} = \frac{1}{2} K s^2 + d s (\rho_1^2 - \rho_2^2)
\;\;\;,
          \end{equation}
where $K$ is the elastic constant and $d$ is another constant that
describes the strength of the lowest order coupling term allowed by
symmetry. Note that the sign of this coupling is opposite for the two
order parameters, because the stripes run in orthogonal directions (we
have picked axes for the tetragonal distortion along the $(110)$ and
$(1\bar10)$ directions).

          Now imagine that we have a homogeneous situation, with $\rho_1
\ne 0$, and $\rho_2 = 0$, so that only one of the
          two kinds of stripes is present. We can determine the magnitude
of the order parameter $\rho_1$, by looking for the minimum in the
          free energy. In the case where the strain is free to relax, we
have from $\partial {\cal F}_{s} /\partial s = 0$ that
          \begin{equation}
          s = -\frac{d \rho_1^2}{K} \;\;\;.
          \end{equation}
          As expected, a shear strain will accompany the stripe
transition (c.f. the structural transition to a monoclinic COI phase
described earlier \cite{Ritter}).
           Substituting this back into the free energy, which is now just
a function of the single variable $\rho_1$, we have
           \begin{equation}
           {\cal F}{(\rho , s(\rho))} = {\cal F_{\rho}} - \frac{d^2}{2K}
\rho_1^4 \;\;\;,
           \end{equation}
           so that the strain relaxation makes the fourth-order
coefficient more {\em negative} in the reduced free energy. 
           This means that a transition that was already first order is
made more strongly so such that $T_o$ is raised. It is also possible
           that initially $b>0$, but that $b-2d^2/K <0$, so that a
second-order transition --- if the strain were clamped ($s=0$) --- can be
driven
           first order if the strain relaxes. Even more dramatically, if
$b$ is sufficiently large and positive, the transition might in fact only
occur at all if the strain is allowed to relax. 
           Such phenomena are quite common in ferro-electric and
ferro-elastic oxides \cite{ferroelectrics}.

          In principle, we should also have a term ${\cal F}_m$,
describing the magnetic free energy. In order not to complicate the
          discussion more than necessary, we shall not include it
explicitly. Instead, we assume that the dominant
          terms are ${\cal F}_{\rho}$, and ${\cal F}_s$, and assert that
(metallic) ferromagnetism will appear as the
          low temperature phase if the COI is suppressed --- i.e.  the
ferromagnetism is slave to the {\em absence} of charge order.

In the manganites there is now a wide body of evidence to demonstrate that
clamping the strain can produce pronounced effects. For example, as
described earlier Section 3.2, compositions in which the equilibrium
strain relaxed phase is the COI tend to adopt the FMM phase if the strain
is clamped in the right way, either by imposed external stresses, or by
{\em internally} generated stresses in multi-domain samples --- such as
micron-sized grains in free powders \cite{Radcondmat}. 

          Strain-induced interactions are long-range, and can lead to
pronounced kinetic effects, familiar in the martensites
          \cite{martensite}.
          Since the shear strain induced by two orthogonal striped
domains is of opposite sign, the strain cancels at long
          distances such that the arrangement shown in Fig. 2a is favoured
by nucleation in a bulk or clamped environment. However the strain fields
are incompatible unless the
          two domains meet in a macroscopically strain-free fashion as
shown in Fig. 2b. Either way there is the possibility of untransformed
material between the two domains, which if sufficiently large we may
expect to become a ferromagnetic metal. 
          \begin{figure}
          \centerline{\includegraphics[width=3in]{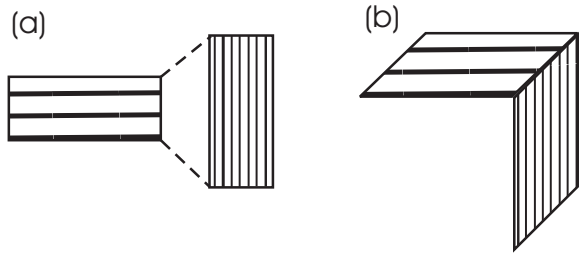}}
          \caption{\label{fig2} (a) Frustrated and (b)  non-frustrated
``$90^\circ$'' twin boundaries between stripe domains.
          Arrangements like (a) guarantee a macroscopic
          amount of untransformed material between the domains, whereas in
(b) such material may only be present at the interface. In the manganites,
any
          macroscopic amount of material that fails to transform into the
COI phase will likely become a FMM.}
          \end{figure}

          \subsection{Twin boundary between striped domains}

          In a $90^\circ$ twin boundary (Fig. 2b), the interface between
the two
different stripe domains bisects the $90^\circ$ angle between the domain
          orientations. Such a configuration potentially allows
macroscopic strain relaxation for both domains (Fig. 2b). If there are
indeed domains
          in a sample, such twins have the lowest strain energy.
          However, the domain boundary cannot
          be atomically abrupt, especially because we do not expect the
different order parameters for each domain to overlap $(b_{12} \gg 0)$; 
neither, equally, do
          we expect these order parameters to jump abruptly to zero. More
likely is the situation sketched in Fig. 3a, where
          the two order parameters decay over a length scale $\xi_o$ which
is typically larger than a lattice constant (and even larger
          for smaller amplitude charge order modulations). If the
coherence length $\xi_o$ is large enough, then it is clearly
          possible that the twin boundary itself is locally a ferromagnetic metal. Synchrotron x-ray diffraction
evidence consistent with this scenario is found in single crystal $La_{5/8-y}Pr_{y}Ca_{3/8}MnO_3$ \cite{Kirynew},
but not, for example, in $(Pr,Sr,Ca)MnO_{3}$ \cite{Kiryprivcomm} implying that the proposed scenario is not
universal. 

          \begin{figure}[h]
          \centerline{\includegraphics[width=3in]{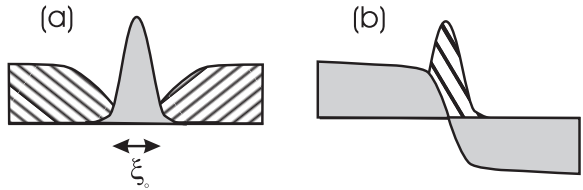}}
          \caption{\label{fig3} Schematic picture of charge-order
parameter (striped) and magnetisation (grey)
          at (a) a twin boundary and at (b)
          a magnetic domain wall, in situations where the COI and FMM
phases are close in energy.}
          \end{figure}

          \subsection{Domain wall in the FMM}
          Competition between COI and FMM could also appear when the
ferromagnetic phase is the stable ground state if we impose
          a magnetic domain wall, as in Fig. 3b. Usually
magnetisation decays near a boundary over a long length scale
          determined by the (Heisenberg) exchange parameter and the
magnetic anisotropy. One would therefore expect a negligibly small
reduction in electronic bandwidth. If however, the suppression of the
magnetisation
          at the FM domain wall were to lead to local charge order
\cite{NDMNaV}, this
will produce a magnetic weak link akin to what is observed in
          {\em fabricated} magnetic tunnel junctions. Such effects may be
responsible for the surprisingly large electrical resistance
\cite{mathur_jap}
          of albeit apparently wide \cite{Lloyd} magnetic domain walls in
thin films. 

          \subsection{Incommensurate striped and magnetic phases}
          In the case of the twin boundary and magnetic domain walls just
described, we imagined that these configurations were
          imposed by external conditions. There is a further possibility,
however, that such configurations may not be thus externally
          (or kinetically) imposed, but might also form long-period
thermodynamic phases.  To see how this might come about, we
          should understand how in the Landau theory we might incorporate
the chemical tendency to match the period of the COI phases
          to the nominal Mn(III)/Mn(IV) ratio. Within the weak coupling
approach of the Landau theory, one can allow for this by
          generalising the Landau theory to a complex order
parameter \cite{mcmillan}
          \begin{equation}
          \label{psi}
          \psi_i = \rho_i ({\bf r}) e^{i({\bf Q}_i \cdot {\bf r} +
\phi_i({\bf r}))} \;\;\;,
          \end{equation}
          where the physical charge density modulation is $|\psi_i({\bf
r})|$. If we choose the period ${\bf Q}_i$ to be commensurate
          with the lattice ${\bf Q}_i = {\bf G}_i/n$, with $n$ an integer,
and ${\bf G}_i$ a reciprocal lattice vector of the parent phase,
          then we can incorporate the possibility of structures with
incommensurate periodicities by the phase variable $\phi({\bf r})$.
          New terms in the free energy dependent upon the phase can have
the following form
          \begin{equation}
          {\cal F}_{\phi} = \xi_o^2 |(\nabla -iq)\psi_i|^2 - g \Re (\psi^n
e^{-i{\bf G}_i \cdot {\bf r}}) + ...
          \end{equation}
          The first term represents the preference for an incommensurate
phase (where $q$ changes with concentration), and the second
          term is a commensurability (Umklapp) term that lowers the energy
when the density wave lines up appropriately with the
          underlying lattice ($g$ is a constant and the operator ${\Re}$
returns the real part of its argument).

          When the amplitude $\rho_i$ is stiff, these terms in the free
energy can be simplified to give
          \begin{equation}
          \label{fphi}
          {\cal F}_{\phi} \approx \xi_o^2 \rho_i^2 |\nabla \phi - q|^2 - g
\rho_i^n \cos(n\phi_i({\bf r}))  \;\;\;.
          \end{equation}
          Here, we assume that the amplitude $\rho_i(T)$ is already fixed
by the minimisation of (2) and (3). The minimisation of (\ref{fphi})
produces the well-known sine-Gordon equation,
          and solutions which are solitons. Without discussing the
details, notice that if the first term in (\ref{fphi}) were dominant, then
          the phase will tend to ``wind'' with gradient $q$, whereas the
second term suppresses the phase slips and prefers the
          commensurate phases $\phi= 2 \pi j / n$. Arrays of solitons
(``discommensurations'') arise at the minimum of the
          free energy intermediate between these limits.
          Furthermore, note that for $n > 2$, the last term is of higher
order in $\rho_i(T)$ than the first, so that there is generically
          the possibility of a transition from a commensurate to an
incommensurate state with increasing $T$ (and reducing $\rho_i$),
          whereas no such transition is expected for $n=2$. This may help
to account for the great stability of
          ``1/2'' type ordering in the manganites, which exists for a wide
range
          of dopings. Incommensurate states are seen near $x=2/3$,
however, and ``frozen-in'' discommensurations are sometimes visible
          in electron micrographs at low temperature in this region of
doping \cite{Chenprivcomm}.

          That the phase and amplitude can be so cleanly separated is an
approximation justified only in the weak
          coupling limit when $\rho_i \ll 1$, and $\xi_o |{\bf Q}_i| \gg
1$.  This is probably a poor approximation for the large amplitude
distortions
          common to the manganites, in which case other potential long
period phases may be favoured by the competition between the
          two principal order parameters, modulated by the imposed
periodicity $q$ (note that $q$ acts like --- and in fact couples to ---
          a lattice strain). One speculation is the possibility of a novel
incommensurate magnetic/CO phase, that can be made by
          a periodic arrangement of either of the configurations shown in
Fig. \ref{fig3}. It is therefore possible that these
          types of configurations may not exist solely because they are
          externally imposed: they may be thermodynamically stable in
their own right.

          \section{Manipulation of self-organised structures}

          We remarked near the end of Section 3.2 that it is not the {\em
coexistence} of different phases in the manganites that is remarkable,
but rather the fact that the
          proportions of these phases can readily {\em adjust} to small
external forces. It seems therefore that the existing components
          are indeed in local equilibrium, and that their coexistence is
not necessarily enforced because of strong hysteresis or pinning.
          The ideas framed in the Landau theory of the previous section
allow for such a delicate balance, because of the
          pervasive effects of long-range strain fields.

          Two (and perhaps more) broad possibilities allow for
quasi-equilibrium coexistence. One is that the domain
          configurations are topologically frozen in. This is
understandable because rearrangements that would remove large domains are
kinetically disallowed.  Nevertheless, local readjustments could easily be
made --- for example to the proportion of the FMM and COI phases in a
frustrated configuration like the one sketched in Fig. 2a.
Another class of possibilities is that one has not just the COI and FMM as
stable thermodynamic phases, but in
          addition a host of incommensurate structures built out of
periodic arrangements of these two basic subunits. This type of
          situation is certainly favoured by the natural tendency to
incommensurability away from special doping values.

The potential range of self-organising structures in the manganites
suggests that it should be possible to produce patterns in a controlled
manner --- in principle for the purpose of information storage. Strain
templating is likely to be an effective means of achieving this because
strain is so important in these materials --- as discussed throughout this
article. The possibility of x-ray lithography \cite{xray} and the use of
currents \cite{currents} to pattern domains has already been suggested.

\section{Conclusions}

The mesoscale complexity that can arise in the manganites is reminiscent
of the complexity that arises in the high temperature superconducting
cuprates, or even organic systems. Because the crystal structure of a
manganite interacts strongly with the corresponding magnetic and
electronic structures, any {\em imposed} physical discontinuities in
actual samples will likely create or destroy the delicate phase
balances that we have described in this article. The challenge is
therefore to better demonstrate the {\em intrinsic} nature of coexistence
in the manganites, and to exploit this coexistence to produce
self-organised structures over mesoscopic length scales.

\section*{Acknowledgements}

We thank for helpful discussions C.J. Adkins, J.P. Attfield, L. Brey, M.J. 
Calder\'{o}n, C.H. Chen, S.-W. Cheong, Cz. Kapusta, P. Levy, A.J. Millis,
D.V. Sheptyakov and B.B. van Aken. We also wish to thank M.G. Blamire,
J.E. Evetts, M.-H. Jo, S.J. Lloyd, J.C. Loudon, P.A. Midgley and N.K. Todd
for support and helpful discussions. We are grateful for the support of
INTAS, the UK EPSRC, the Royal Society and Churchill College, Cambridge.

\end{document}